# Thermal Interface Conductance between Aluminum and Silicon by Molecular Dynamics Simulations

Nuo Yang,[1,2] Tengfei Luo,[3] Keivan Esfarjani,[4] Asegun Henry,[5a,b] Zhiting Tian,[2] Junichiro Shiomi,[6] Yann Chalopin,[7] Baowen Li,[1,8,9*] Gang Chen[2*]

[1]Center for Phononics and Thermal Energy Science, School of Physical Science and Engineering, Tongji University, 200092 Shanghai, China

[2]Department of Mechanical Engineering, Massachusetts Institute Technology, Cambridge, 02139 Massachusetts, USA

[3]Department of Aerospace and Mechanical Engineering, University of Notre Dame, Notre Dame, IN 46556, USA

[4]Department of Mechanical Engineering, Rutgers University, Piscataway, 08854 NJ, USA

[5a]George W. Woodruff School of Mechanical Engineering, Georgia Institute of Technology, Atlanta, GA 30332, USA

[5b]School of Materials Science and Engineering, Georgia Institute of Technology, Atlanta, GA 30332, USA

[6]Department of Mechanical Engineering, The University of Tokyo, 7-3-1 Hongo, Bunkyo-ku, Tokyo, 113-8656, Japan.

[7]Laboratoire d'Energétique Moléculaire et Macroscopique, CNRS UPR 288, Ecole Centrale Paris, Chatenay-Malabry, France

[8]Department of Physics and Centre for Computational Science and Engineering, National University of Singapore, 117456 Singapore, Republic of Singapore

[9]NUS Graduate School for Integrative Sciences and Engineering, National University of Singapore, 117456 Singapore, Republic of Singapore

Corresponding Author: * E-mail: (GC) gchen2@mit.edu; (BL) phylibw@nus.edu.sg

## Abstract

The thermal interface conductance between Al and Si was simulated by a non-equilibrium molecular dynamics method. In the simulations, the coupling between electrons and phonons in Al are considered by using a stochastic force. The results show the size dependence of the interface thermal conductance and the effect of electron-phonon coupling on the interface thermal conductance. To understand the mechanism of interface resistance, the vibration power spectra are calculated. We find that the atomic level disorder near the interface is an important aspect of interfacial phonon transport, which leads to a modification of the phonon states near the interface. There, the vibrational spectrum near the interface greatly differs from the bulk. This change in the vibrational spectrum affects the results predicted by AMM and DMM theories and indicates new physics is involved with phonon transport across interfaces.

## Introduction

Thermal transport across interfaces is an important issue for microelectronics, photonics, and thermoelectric devices and has been studied both experimentally and theoretically in the past.[1-8] Although recent experiments using pump-probe methods are performed on metal/dielectric (or semiconductor) interfaces, most simulations are for dielectric/dielectric interfaces for which only lattice vibrations are involved. There are three channels for heat transport across metal–semiconductor interfaces: 1) direct phonon-phonon coupling between phonons in the metal and phonons in the semiconductor; 2) direct coupling between electrons of the metal and phonons of the semiconductor; 3) indirect coupling between electrons in the metal and phonons in the semiconductor through electron-phonon interactions on the metal side, close to the interface, and subsequent phonon-phonon interactions across the interface.[9] A few treatments in the past have considered the effects of electron-phonon coupling in metal on the thermal boundary resistance between a metal and a dielectric.[10-14] Sergeev suggested that the second channel is significant for conductors with strong electron-phonon coupling, or for an interface with high phonon reflectivity.[11] Majumdar and Reddy showed that the third channel cannot be ignored and could even play a dominant role at high temperatures.[12] Mahan presented a theory of heat flow between the electrons in a metal and the phonons in an ionic crystal, which interact through the surface charges caused by the image potential.[14]

The nonlinear forces between atoms in heat transport across interfaces make predictions intractable from an analytical point of view. Since models based on diffuse mismatch (DMM) and acoustic mismatch (AMM) do not work well above cryogenic temperatures, molecular dynamics (MD) simulations have become the method of choice for predicting interface thermal conductance over the last decade.[1,2] Using the non-equilibrium simulation, Landry and McGaughey show that the interface thermal conductance of Si/Ge increases with increasing temperature from 0.33 $GWm^{-1}K^{-2}$ (300 K) to 0.67 $GWm^{-1}K^{-2}$ (1000 K).[15] Chalopin et al., on the other hand, used an equilibrium MD method to study thermal conductance at Si/Ge interfaces,[6] and both the period thicknesses dependence and the temperature dependence of thermal conductance in Si/Ge superlattices were presented.

However, there are few MD simulation studies of the interface conductance of metal–semiconductor interfaces, because of the inherent difficulty associated with including contributions of both electrons and phonons to heat transport in MD simulation.[16] Using MD simulations, Cruz *et al.* calculated the thermal interface conductance of Au/Si at 300 K,[17] and the value of 188 $MW/m^2$-K is in good agreement with the measurement results with values ranging between 133 and 182 $MW/m^2$-K.[18] Including electron-phonon couplings in MD simulation using the method described by Duffy and Rutherford,[19] Wang *et al.* calculated interfacial thermal conductance of Cu/Si with values around 400 $MW/m^2$-K, which is in better agreement with experimental data compared to those without electron-phonon couplings in MD (~450 $MW/m^2$-K).[20]

In this letter, we calculated the thermal interface conductance of an Al/Si interface using non-equilibrium MD (NEMD) simulations, considering both the first and the third channels' contribution to heat transfer across interfaces. In section 2, we describe the structure of the model followed by the description of MD simulation method, and we present the results of the interface conductance and related discussions in section 3.

## Molecular dynamics simulation

The second nearest-neighbor modified embedded atom method (2NN MEAM) interatomic potential, which is based on density functional theory, was used to describe the atomic interactions. The detailed parameterization of the potential is presented by Lee et al..[21] The 2NN MEAM was chosen because it has been applied to a wide range of elements including body-centered cubic (bcc), face-centered cubic (fcc), hexagonal close-packed (hcp) metals, manganese, and diamond-structured covalent bonding elements, and their alloys using one common formalism.[17,21-23] Thus, a single potential can be used to describe both of the materials of interest - silicon and aluminum. This greatly simplifies the description between the two materials at the interface region, compared to treating them with potentials of different functional forms. The 2NN MEAM has a relatively long range interaction, and the energy of free surfaces can be described more accurately than standard MEAM potential and the widely used Stillinger-Weber potential,[24] which only includes interactions within the first neighbor shell, and consequently fails in describing surface reconstruction and therefore may become inaccurate near interfaces. The velocity Verlet algorithm is used to integrate the discretized differential equations of motions.

Electron-phonon scattering of each atom (i) in the metal was introduced into the simulations by describing the energy loss via a friction term with coefficient ($\gamma_i$) and the energy gain from electrons via a stochastic force term $\tilde{F}(t)$ with random magnitude and orientation:[19]

$$m_i \frac{d^2 r_i}{dt^2} = -\nabla_{r_i} V(r_i) - \gamma_i \frac{dr_i}{dt} + \tilde{F}(t) \quad (1)$$

$$\gamma_i = \frac{g m_i \Delta V}{3 N k_B} \quad (2)$$

where N is the total number of atoms in the cell with constant electronic temperature, $\Delta V$ is the cell volume, and g is the electron-phonon coupling constant. $g_{Al}$ is set as $2.45\times 10^{17}$ $Wm^{-3}K^{-1}$ for 300 K, which is obtained from ab-initio electronic structure calculations using density functional theory.[25] For equilibrium systems the magnitude of the stochastic force is related to the friction coefficient by the fluctuation–dissipation theorem and the energy exchange drives the atomic system to the temperature of the electronic subsystem ($T_e$) as:

$$\left\langle \tilde{F}(t) \right\rangle = 0, \qquad \left\langle \tilde{F}(t') \cdot \tilde{F}(t) \right\rangle = 2k_B T_e \gamma \delta(t'-t) \tag{3}$$

Since electrons reach equilibrium much faster than phonons in a simulation cell of several nanometers, it is reasonable to assume $T_e$ as a constant which is reasonable for nanoscale cell.

In this paper, we study the thermal interface conductance of Al/Si along [100] direction. As shown in Fig.1 (a), different interface distances lead to changes of interfacial energy calculated based on 2NN MEAM. The interface distance (defined in Fig.1 (b)) is chosen as 0.26 nm which corresponds to the minimum of interfacial energy density. Periodic boundary conditions are imposed in both two transverse dimensions (x and y) and the longitudinal direction (z). The mass of Si atoms is set randomly in accordance with the natural isotopic abundance (92% $^{28}Si$, 5% $^{29}Si$ and 3% $^{30}Si$).

The lattice constants of Al and Si are 0.4047 ($a_{Al}$) and 0.5431 ($a_{Si}$) nm, respectively, which leads to a lattice mismatch of the cross section. This stress and strain is a general problem when simulating two different lattices with periodic boundary conditions. In order to reduce stress, a paring with the smallest value of the percentage of cross section mismatch (defined as the difference between the cross section area of Al and area of Si over the larger one) is needed. When paring $4n\times 4n$ (n=1,2,3,…) unit cells of Al and $3n\times 3n$ unit cells of Si, the percentage of cross section mismatch is 0.12% which smaller than other pairings within the computationally affordable range of n. Al 8×8 pairing up with Si 6×6 unit cells is used in our simulation. However, there is still a subtle lattice mismatch. To test the effect of this interfacial strain, two values, 10.5nm$^2$ ($8a_{Al} \times 8a_{Al}$) and 10.6nm$^2$ ($6a_{Si} \times 6a_{Si}$), were used as cross section areas of the simulation cell. Results show that the difference of values of thermal conductance using these two simulation cells is 5%, which indicates that the stress and strain effect on our simulation results is not very significant.

Before calculating the thermal interface conductance, the structure was fully relaxed. The atoms moved freely to release the interfacial energy for more than 20 ps in a microcanonical ensemble (NVE) process. The atoms were then thermalized to 300 K during a 20 ps interval by Nosé-Hoover method,[26] and finally kept there for 60 ps before the data were collected. MD simulation time step, Δt, is chosen as 1 fs.

The NEMD method generates a temperature gradient across the simulation cell along the direction perpendicular to the interface by introducing heat baths at the two sides of the interface, which mimic the conditions of macroscopic experiments. In order to establish a temperature gradient along the longitudinal direction, a few atomic layers are controlled by heat baths with temperatures $T_L$ and $T_H$ imposed by using the Nosé-Hoover thermostat.[26] Simulations are carried out long enough such that the system reaches a steady state. Then, the kinetic temperature ($T = \sum m_i v_i^2/2$) at each perpendicular atomic plane (x-y plane) and the heat flux in each thermal bath are averaged over nanoseconds. In Fig. 2 we show the typical time-averaged temperature profile. In the inner regions, the temperature profile is fitted with a linear function. The temperature profiles close to the interface are nonlinear because of the atomic level disorder near the interface.

The interface thermal conductance (G) and heat flux (J) are calculated as

$$G = \frac{J}{A \cdot \Delta T} \tag{4}$$

where ΔT is the interface temperature difference and A is the cross section area. The heat transferred across the interface can be calculated from

$$J = \frac{1}{N_t} \sum_{i=1}^{N_t} \frac{\Delta \varepsilon_i}{2 \Delta t} \tag{5}$$

where Δε is the energy added to/removed from each heat bath for each time step Δt. We use a combination of time and ensemble sampling to obtain better average statistics. The results represent averages from 8 independent simulations with different initial conditions. Each case runs longer than 1 ns after the system reached the steady state.

## Results and discussions

To demonstrate the validity of a classical MD simulation with electron-phonon couplings, we calculated the phonon thermal conductivity of Al using Green-Kubo methods. The values of phonon thermal conductivity of Al are 16.8, 16.3 and 16.4, when the sizes of simulation cell are 4×4×4, 6×6×6, 8×8×8, respectively. As shown in Table I, our MD results are compared to theoretical predictions[12,27] and simulation result from others.[28] Our results show that the phonon thermal conductivity of Al is decreased when considering electron and phonon scatterings. The thermal conductivity is decreased to 16.0 $Wm^{-1}K^{-1}$ when the value of $g_{Al}$ is an artificial value, 2.45 $\times 10^{18}$ $Wm^{-3}K^{-1}$.

The temperature profile of Al/Si structure is shown in Fig.2. The value of interface thermal conductance by NEMD depends on how one defines the temperature drop (ΔT) at the interface, shown in Eq.(4). This choice is somewhat arbitrary and therefore difficult to deal with due to discontinuous temperature variations close to the boundary region. In general, the interface temperature drop could be defined as the temperature difference at the interface of Al and Si based on linear fitting to the data further away from the interface, by extrapolating the linear temperature gradient in each material to the interface, named as $\Delta T_{fit}$ (shown in Fig.2). Another method to define the interface temperature difference,[29] $\Delta T'$, is to use just the temperature drop across the two interfacial layers. The definition of $\Delta T_{fit}$ includes both interfacial temperature drop and temperature drops inside each material caused by atomic level disorder/mixing near the interface (shown in Fig.1(c)). There is a difference between the temperature of the electrons and the temperature of lattice in the metal. As the temperature was based on $T_e$ in the pump-probe experiment,[30] we also use $\Delta T_{fit,e}$ (shown in Fig.2) to calculated the thermal interface conductance.

It has been known that NEMD simulations suffer from not only artificial propagation of ballistic phonons from the thermostated region, which have an unknown distribution, but also convergence with respect to size.[15] The size effect comes from internal reflections of phonons from the boundaries and insufficient phonon-phonon scattering in the simulation cell, which limit their mean free path compared to the true bulk system.[26,31-33] For this reason, we have performed NEMD simulations with several different lengths using the same cross section and with the same temperature. Different from the bulk resistances extracted from the slope, the interface resistances are extracted from the discontinuity in the linearized temperature profile at the interface. Bulk values of the thermal resistance can be obtained from the calculated resistances with finite size and then extrapolated to infinite size according to[34]

$$\rho(\infty) = \rho(L) + C/L \tag{6}$$

where C is a constant, and L is the length of the simulation cell. Similarly, we use the same approach to extrapolate the interfacial thermal resistance to infinite length, as

$$1/G(\infty) = 1/G(L) + C/L \tag{7}$$

Our results show that the interfacial thermal resistance increases with increasing cell length (Fig. 3). Close to the interface, the atomic arrangement is disordered and the interatomic distances are slightly contracted or expanded; while the atoms furthest away from the interface maintain their respective lattice positions. There are obvious differences in each region's power spectra (shown in Fig.4 (a)), which are responsible for a bigger thermal interface resistance. For a shorter supercell, all atoms are close to the interface and their power spectra in the two regions converge which corresponds to a smaller thermal interface resistance.

The linear fits in Fig. 3 can be used to estimate the thermal conductance in the limit of infinite length. As $L \to \infty$, the value of thermal conductance is calculated as 0.47 $GWm^{-2}K^{-1}$ when there is no electron-phonon coupling in Al, and 0.45 $GWm^{-2}K^{-1}$ ($\Delta T_{fit}$ used) and 0.36 $GWm^{-2}K^{-1}$ ($\Delta T_{fit,e}$ used) when electron-phonon couplings are included. Using transient thermo-reflectance method, the thermal interface conductance value between Al and Si are measured to be 0.35 $GWm^{-2}K^{-1}$ [35] and 0.22 $GWm^{-2}K^{-1}$ [36] at 300 K. Hopkins *et al.* showed that the thermal boundary conductance decreases as Si surface roughness increases,[37] where the highest values the thermal interface conductance between Al and Si is close to 0.2 $GWm^{-2}K^{-1}$ at 300 K. By using DMM and taking into account the full phonon dispersion relationship over the entire Brillouin zone, Reddy *et al.* calculated the thermal interface conductance values of Al/Si as 0.27 $GWm^{-2}K^{-1}$ at 300 K.[13]

Our simulation results are closer to these measured values and theoretical predictions. The interface structure is simulated with disordered structure and isotopic impurities. Collins *et al.*

show imperfect surface structures could be an important factor in thermal interface conductance reduction.[38] When $\Delta T_{fit,e}$ is considered, the value is closer to the experimental value.

In making these comparisons, we should also keep in mind that G is calculated based on the extrapolated temperature difference ($\Delta T_{fit,e}$), which is about five times larger than the interfacial temperature drop, ΔT’, (shown in Fig. 2) and that, in the pump-probe experiment, the temperature was based on that of electrons in the metal side ($T_e$).[30]

To investigate the effect of electron-phonon coupling on the interface thermal conductance, we compared the system with and without electron-phonon coupling. Our MD results show that the interface thermal conductance is decreased by about 4% with the effect of electron-phonon coupling at room temperature (shown in Fig. 3), due to the third channel (mentioned in the introduction) of heat transport across Al/Si interfaces. The electron-phonon coupling adds an extra resistance to the system if the energy must be relaxed from electrons in Al to phonons in Al and then from phonons in Al to phonons in Si forming a series of thermal resistances. That is, electrons have two effects: one is to increase resistance due to electron-phonon coupling; the other is electron-phonon coupling also influence phonons which impact phonon-phonon interface resistances.

As proposed by Majumdar and Reddy,[12] the resistance of electron-phonon in the metal and the resistance of interfacial phonon-phonon are in series, $1/G = 1/G_{ep} + 1/G_{pp}$. We also use this equation to calculate. $G_{ep} = \sqrt{g_{Al}\kappa_{ph,Al}} = 2.0$ $GWm^{-2}K^{-1}$ and $G_{pp} = 0.46$ $GWm^{-2}K^{-1}$, then $G = 0.37$ $GWm^{-2}K^{-1}$, which is close to our MD results.

To understand the mechanism of interface resistance, the vibration power spectra are calculated and displayed in Fig.4 (a), where the grouping of atoms is shown in Fig.1 (a). The phonon power spectral density (PSD) describes the power carried by the phonon per unit frequency. A high PSD value for a phonon with frequency $f$ means that there are more states occupied by it. A zero means there is no such a phonon with $f$ exist in system. The phonon power spectrum analysis provides a noninvasive quantitative means of assessing the power carried by phonons in a system. The spectra are obtained by Fourier transforming of velocities. In the spectra of Si far from interface (Group A), it shows that the frequencies of phonons in Si using 2NN MEAM are reproduced with reasonable accuracy, which is much better than results using MEAM[39] where the frequency of optical modes is too high. The spectrum of Si (Al) at the interface, Group B (C), is broader compared to the spectra of Si (Al) far from the interface, Group A (D). It shows the atomic level disorder near the interface increase diffusive scatterings, which contribute the interface resistance and nonlinear temperature profile close to interface in Fig.2. In the case of large interfacial thermal resistance, the broader power spectra can add new anharmonic channels through which phonons can be transmitted to the other side. However, in the acoustically well-matched case, it does not help. The broadening in spectrum reduces in the low-frequency region where there is good acoustic matching, and increases high frequency states which contribute very little to the transfer of heat.

It is intriguing that the Si (Al) atoms close to the interface vibrate at higher frequencies than they do far from the interface, where around 17% (22%) of phonons have their frequency higher than the highest possible vibration of Si (Al). This translates to new states that do not exist in the bulk material.

After the system is cooled down to 0.01 K by a damping process, the force constant of each atom, i, is calculated as:

$$\Gamma_i = -\frac{1}{6}\sum_{d=1,2,...,6}\Delta F_{i,d}/\Delta r_{i,d} \qquad (8)$$

d=1,2,…,6 correspond to $\Delta r$ in six directions, ±x ±y and ±z. The lattice structure of Al/Si interface (Fig1(c)) is not ideal crystal lattice. For example, a number of Si atoms near the interface have coordination numbers that deviate from 4, as is the case in the bulk material. There are bonds that can be much shorter or longer than the equilibrium bond length. Depending on the bonds and the anharmonic 2NN MEAM potential, force constants at interface become larger or smaller than the values of bulk (Fig.4 (b)).

Due to stiffer bonds at the interface, the highest possible vibration of Si shifts from about 15 THz to 20 THz, approximately 30% higher. The higher frequency phonons can not participate in the energy transport across interface without phonon scatterings, because such modes only exist close to the interface and can not well-matched with modes far from the interface. The interface atoms have different crystal structure from the bulk due to the force change.

The popular theories, AMM and DMM, are based on the mismatch of frequency spectra of the two materials. This is usually taken as the bulk vibrational spectrum. We find that the vibrational spectrum changes near the interface, which would affect the results predicted by AMM and DMM theories. The atomic level disorder near the interface is an important aspect of interfacial phonon transport, which leads to the interface phonon states greatly differing from the bulk states and introduces new physics in researching the phonon transport across the interfaces. On investigating the interfacial thermal conductance, one needs to consider the interface states that can greatly differ from the bulk states. Despite the well known theories based on the mismatch of the bulk properties, our results highlight an important aspect of interfacial phonon transport that is essential in understanding realistic surfaces.

## Conclusions and outlook

The thermal transfer in Al/Si structure was simulated by NEMD methods with/without considering electron-phonon coupling as a random noise on Al atoms. The value of thermal interface conductance of Al/Si is calculated by extrapolating length to infinity, where the electron-phonon scatterings decrease the interface thermal conductance. The results show that the interface thermal conductance increases with the decreasing of the system length. The nonlinear discontinuity in temperature profile extends a few atomic layers beyond the interface, which could be attributed to the confinement of phonons and the interface stress and strain effects on phonon transport. Through the difference in power spectra, we find that the atomic level disorder near the interface is an important aspect of interfacial phonon transport, which greatly modifies phonon states near the interface differ from the bulk states. This change in the vibrational spectrum indicates the presence of non-bulk vibrational modes which appear to be involved with phonon transport across interfaces.

The paper deals with thermal interface conductance between metal and semiconductor. Other forms of thermal interface conductance are given for example in the Refs. [[1,2]]. To obtain more exact results of thermal interface conductance, it is necessary to include the direct coupling between electrons of the metal and phonons of the semiconductor in MD simulation, which is still an open question.[16,20,40]

*Acknowledgements* This work was supported in part by the "Solid State Solar-Thermal Energy Conversion Center ($S^3$TEC), an Energy Frontier Research Center funded by the U.S. Department of Energy, Office of Science, Office of Basic Energy Sciences under Award Number:

DE-SC0001299/DE-FG02-09ER46577 (GC). This work was supported in part by the R-144-000-300-112 from the National University of Singapore (BL). This work is supported in part by the startup fund from Tongji University, Natural Science Foundation of China "11334007" (NY and BL) and "11204216" (NY). NY is sponsored by Shanghai Pujiang program.

Table I Phonon thermal conductivity ($\kappa_{ph}$) of Al and thermal interface conductance (G) of Al/Si at 300 K with different methods

| | MD | Experiments | Theory |
|---|---|---|---|
| $\kappa_{ph,Al}$ [$Wm^{-1}K^{-1}$] | 17.0±0.5[(a)]<br>16.4±0.5[(b)(c)]<br>16.0±0.5[(b)(d)]<br>20.6[(a)28] | - | 10~20[12]<br>10~24[27] |
| $G_{Al/Si}$ [$GWm^{-2}K^{-1}$] | 0.47[(a)(e)]<br>0.45[(b)(e)]<br>0.36[(b)(f)] | 0.35[35]<br>0.2[37]<br>0.22[36] | 0.27(DMM)[13]<br>22.9(AMM)[1]<br>0.37[(c)(g)] |

(a) MD without electron-phonon coupling in Al

(b) MD with electron-phonon coupling in Al

(c) $g_{Al}$ is $2.45\times10^{17}$ $Wm^{-3}K^{-1}$ (Ref.[25])

(d) $g_{Al}$ is $2.45\times10^{18}$ $Wm^{-3}K^{-1}$ (artificial)

(e) $\Delta T_{fit}$ is used

(f) $\Delta T_{fit,e}$ is used

(g) Calculated by $1/G = 1/G_{ep} + 1/G_{pp}$

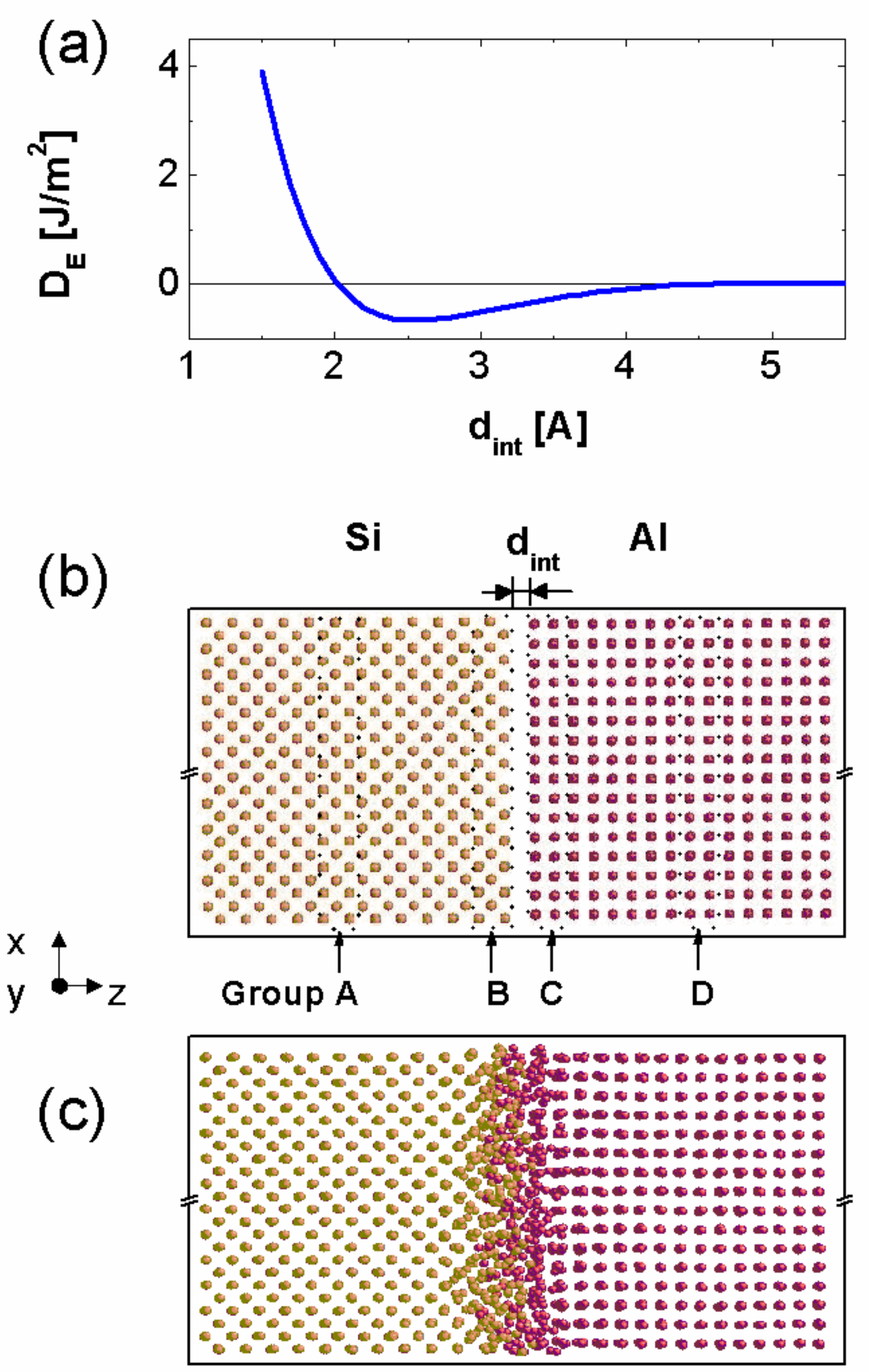

FIG. 1 (a) The dependence of interfacial energy density on the interface distance. The minimum of interface energy density corresponds to the interface distance of 0.26 nm. (b) Simulation cell of Al/Si interface structures before relaxation. The cross section of simulation cell is Al 8×8 / Si 6×6 unit cells$^2$ (3.26×3.26 nm$^2$). The interface distance is defined as $d_{int}$. Periodic boundary conditions are imposed in three dimensions. (c) Al/Si interface structures after relaxation. The atomic level disorders near the interface are obvious.

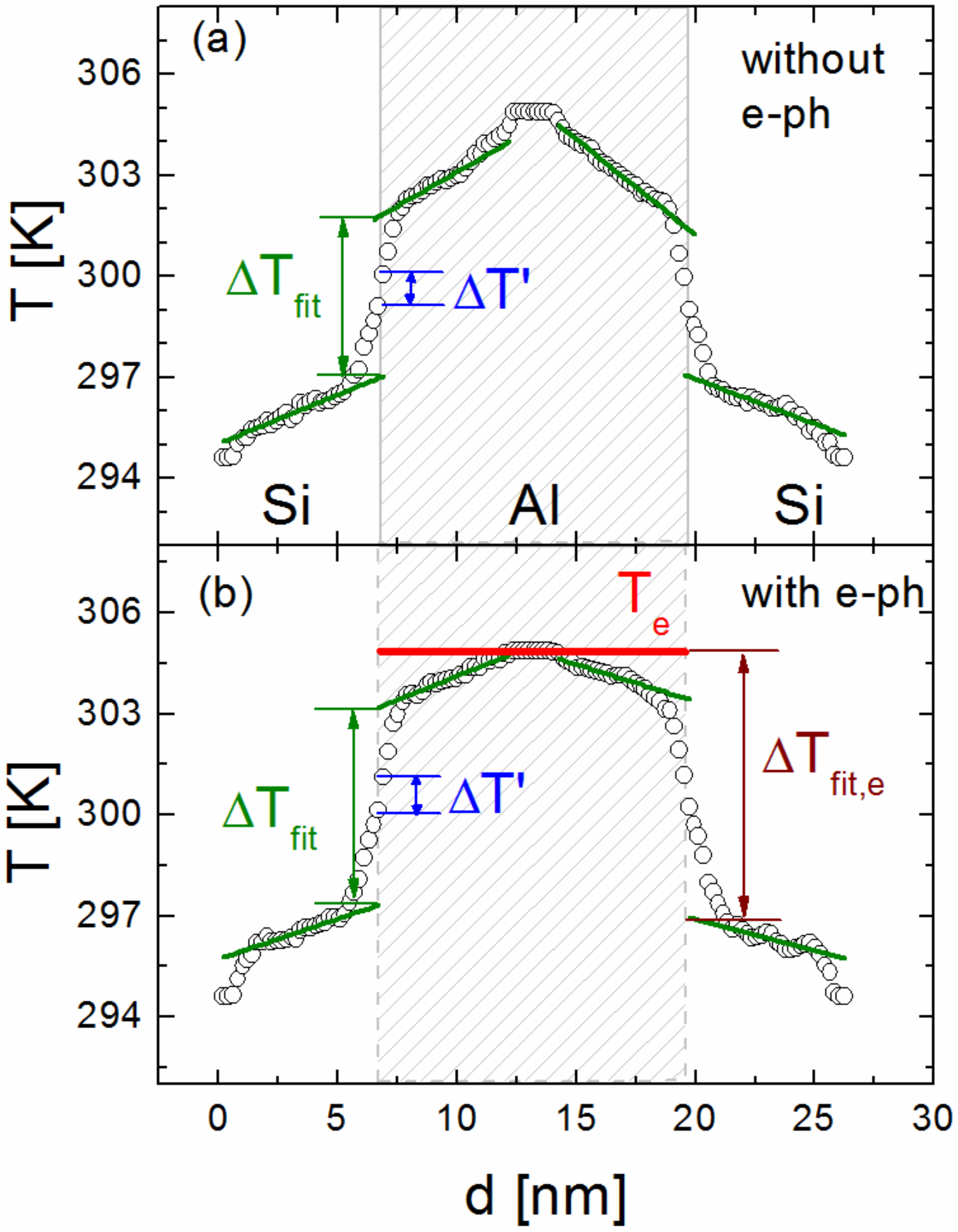

FIG. 2 (a) and (b) are the temperature profile of Al 8x8x32 / Si 6x6x24 units$^3$ without/with considering electron-phonon coupling in Al, respectively. High/Low temperature Nose-Hoover heat bath is applied on several center layers of Al (Si). Temperature gradients with different direction are built along cross-interface direction. There are discontinuous and nonlinear effects close to interfaces, which are caused by the atomic level disorder near the interface. Solid lines are linear fitting curves based on the data further from interfaces. The red line ($T_e$) in (b) shows electron-phonon temperature non-equilibrium in the metal.

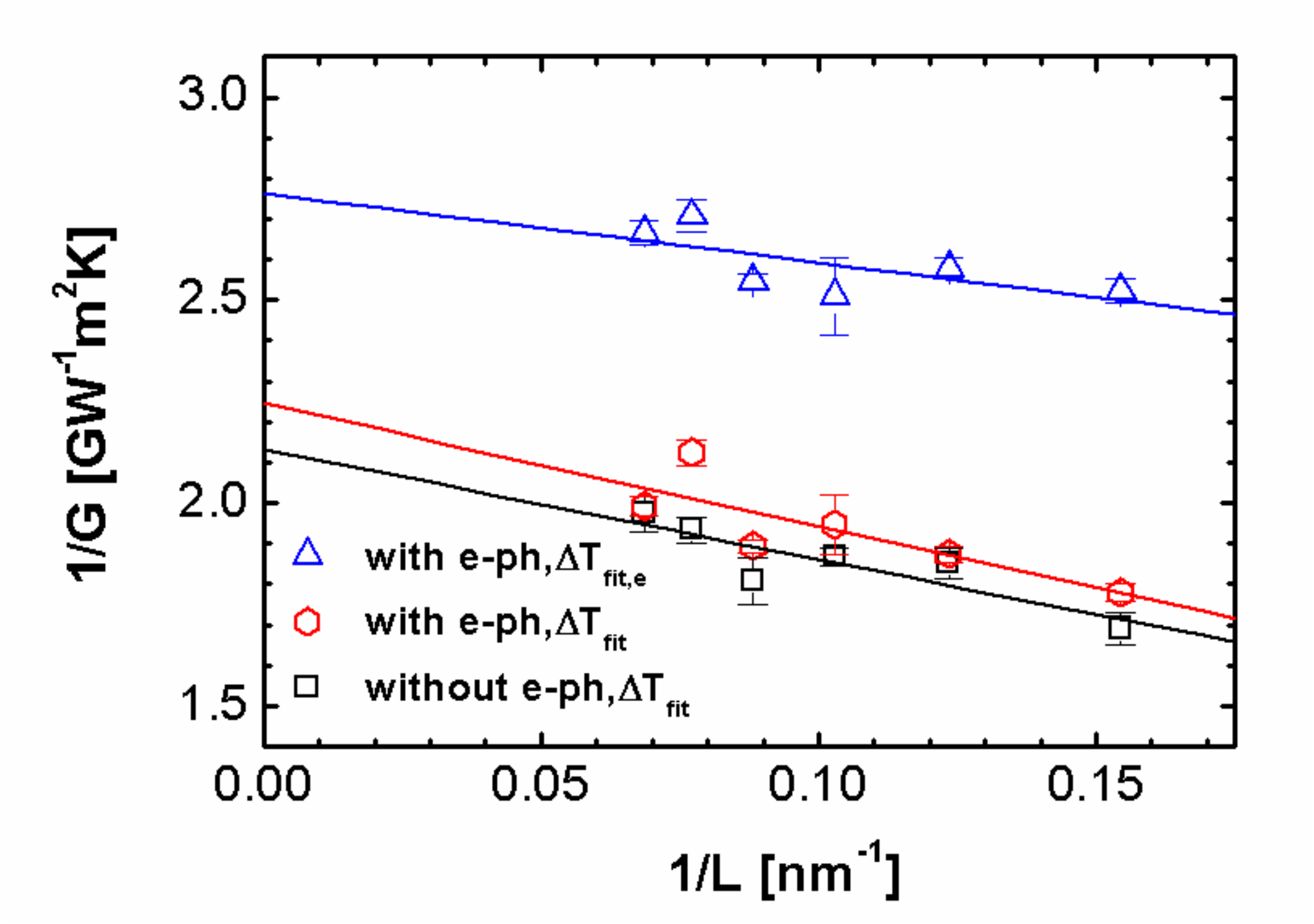

FIG. 3 The size dependence of thermal interface resistance (1/G) on the inverse of length (1/L) at room temperature for Al/Si system by NEMD method. The fitting curves are based on Eq.(7). The blue (red) fitting line correspond to using $\Delta T_{fit,e}$ ($\Delta T_{fit}$) in calculating G with considering electron-phonon couplings on Al atoms. The black fitting line is without electron-phonon couplings.

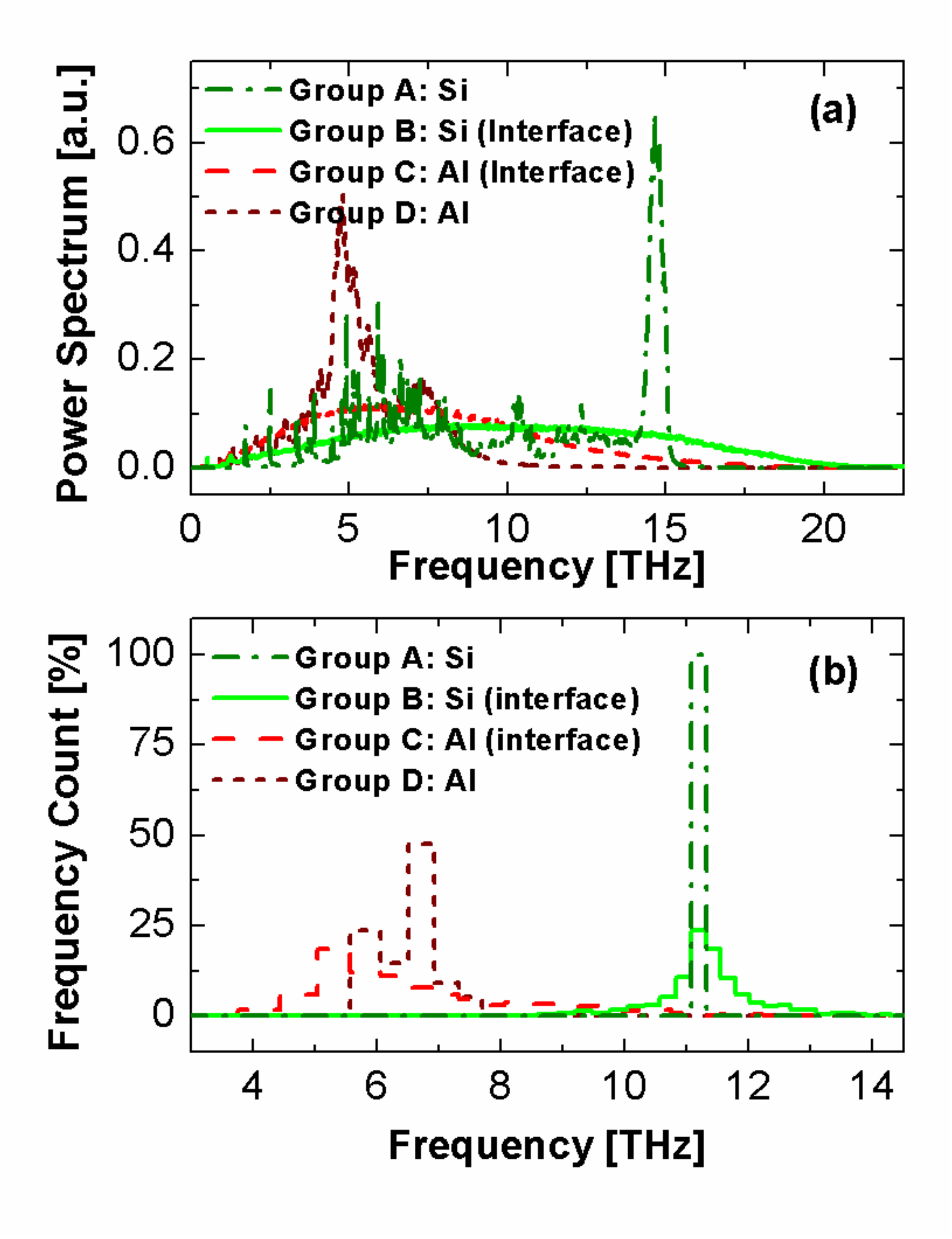

FIG. 4 (a) Normalized power spectra of atoms in Al/Si structure. (b) Frequency count of harmonic oscillator frequency. The harmonic oscillator frequency (f) is calculated based on the force constant ( $\Gamma$ ) as: $f = (\Gamma/m)^{1/2}/(2\pi)$, where m is mass of Al atom or Si atom. The number of atoms in group A and B is 576. The number of atoms in group C and D is 512.